\documentclass[preprint,12pt]{elsarticle}

\usepackage{amssymb}

\usepackage{amsmath}

\newcommand{\define}{\equiv}
\newcommand{\Li}{\text{Li}}
\newcommand{\bin}[2]{\left( \!\! \begin{array}{c} #1 \\ #2 \end{array} \!\! \right)}
\newcommand{\nn}{\nonumber}
\newcommand{\id}{\mathrm{d}}
\newcommand{\floor}[1]{\left \lfloor #1 \right \rfloor}

\usepackage[T2A,T1]{fontenc}
\makeatletter
\def\easycyrsymbol#1{\mathord{\mathchoice
  {\mbox{\fontsize\tf@size\z@\usefont{T2A}{\rmdefault}{m}{n}#1}}
  {\mbox{\fontsize\tf@size\z@\usefont{T2A}{\rmdefault}{m}{n}#1}}
  {\mbox{\fontsize\sf@size\z@\usefont{T2A}{\rmdefault}{m}{n}#1}}
  {\mbox{\fontsize\ssf@size\z@\usefont{T2A}{\rmdefault}{m}{n}#1}}
}}
\makeatother

\newcommand{\sha}{\easycyrsymbol{\cyrsh}}
\newcommand{\shcha}{\easycyrsymbol{\cyrshch}}

\newcommand\stuffle[1]{\mathrel{\overset{\makebox[0pt]{\mbox{\normalfont\tiny\sffamily $#1$}}}{\shcha}}}

\newcounter{bla}

\journal{Computer Physics Communications}

\begin{document}

\begin{frontmatter}

\title{Generalized Polylogarithms in Maple}

\author[a,b]{Hjalte Frellesvig\corref{author}}

\cortext[author] {Corresponding author.\\\textit{E-mail address:} hjalte.frellesvig@kit.edu}
\address[a]{Institute for Theoretical Particle Physics (TTP), Karlsruhe Institute of Technology, Engesserstra{\ss}e 7, D-76128 Karlsruhe, Germany}
\address[b]{Institute of Nuclear and Particle Physics, NCSR ``Demokritos'', Patriarchou Gregoriou E. {\&} Neapoleos 27, Agia Paraskevi, 15310, Greece}

\begin{abstract}

This paper describes generalized polylogarithms, multiple polylogarithms, and multiple zeta values, along with their implementation in Maple 2018. This set of related functions is of interest in high energy physics as well as in number theory. Algorithms for the analytical manipulation and numerical evaluation of these functions are described, along with the way these features are implemented in Maple.

\end{abstract}


\end{frontmatter}


{\bf PROGRAM SUMMARY}

\begin{small}
\noindent
{\em Program Title:} The functions GeneralizedPolylog, MultiPolylog, and MultiZeta. \\


\noindent
{\em Programming language:} \textbf{MAPLE 2018} \\


\noindent
{\em Nature of problem:}\\
Manipulation and numerical evaluation of generalized polylogarithms, multiple polylogarithms, and multiple zeta values. \\

\noindent
{\em Solution method:}\\
The numerics is implemented through the infinite sum that defines the multiple polylogarithm. Outside their convergent region functions are mapped thereto using various identities along with integral level transformations. For the multiple zeta values and for badly convergent cases, the convergence is accelerated using H{\"o}lder convolution. The analytical aspects are implemented in a package, and in the Maple {\tt Transformations} framework for special functions.\\



\end{small}

\section{Introduction}
\label{sec:intro}

Generalized polylogarithms\cite{Goncharov:1998kja, Goncharov:2010jf} (Also known as Goncharov polylogarithms, generalized harmonic polylogarithms, or hyperlogarithms) are a class of functions that frequently appear in analytical expressions for Feynman integrals.

Feynman integrals are mathematical objects that are needed for precise predictions for the results of high energy particle scattering processes, as they take place for instance at the Large Hadron Collider at CERN. (For overviews of the mathematical aspects of Feynman integrals, see e.g. refs. \cite{Smirnov:2012gma, Henn:2014qga, Duhr:2014woa}.) This means that tools for their manipulation and evaluation are of high importance to physicists trying to derive concrete predictions from the theory of elementary particles.

Generalized polylogarithms are a generalization of functions such as the logarithm, the classical (or Euler) polylogarithm, and the harmonic polylogarithm \cite{Remiddi:1999ew}, which all appear as special cases. 

When evaluated at certain special values, generalized polylogarithms reduce to a set of numbers called multiple zeta values \cite{Zagier1994, Broadhurst:1996kc, Blumlein:2009cf}, which are a generalization of the values of the Riemann zeta function evaluated at positive integers. Aside form their appearance in physics, these numbers are also of interest in pure mathematics such as number theory.

Various numerical implementations \cite{Vollinga:2004sn, Maitre:2005uu, Frellesvig:2016ske} of these and related functions exist and are used by the high energy physics community. Additionally various packages for symbolic manipulations of these functions exist \cite{Ablinger:2014rba, Panzer:2014caa, DuhrPolyLogTools}. The Maple 2018 implementation described in this paper is, however, the first general implementation that combines the analytical and numerical aspects into one integrated whole, which aims for broad utility to mathematicians and physicists alike.

The aim of this paper is to describe the three functions the generalized polylogarithm, the multiple polylogarithm, and the multiple zeta values, including all those aspects thereof that are part of the Maple implementation.

In section \ref{sec:DefinitionsAndRelations} we define the various functions, and list a number of their properties and some relations between them. In section \ref{sec:numerics} we define the algorithms used for their numerical implementation. And in section \ref{sec:implementation} we list the various Maple procedures implementing the functions and relations, as they appear in Maple 2018.

\section{Definitions and relations}
\label{sec:DefinitionsAndRelations}

This section contains descriptions of all those features of the polylogarithmic functions that are part of the Maple implementation. Section \ref{sec:definitions} contains the definition of the implemented functions, and section \ref{sec:divergences} describes their divergences. Section \ref{sec:relations} describes a number of relations between the functions, and section \ref{sec:specialvalues} describes some special values. Section \ref{sec:derivatives} describes how to take derivatives of polylogarithmic functions, and finally section \ref{sec:symbols} describes the ``symbol algebra'' sometimes used for their manipulation.

\subsection{Definitions}
\label{sec:definitions}

The Generalized Polylogarithm (GPL) is defined\cite{Goncharov:1998kja, Goncharov:2010jf} recursively, as
\begin{align}
G(a_1, \ldots, a_n; x) \define \int_0^x \frac{dy}{y - a_1} G(a_2, \ldots, a_n; y) \,.
\label{eq:gpldef}
\end{align}
The recursion stops, as $G(;x) \define 1$. When all the $a_i$ indices equal $0$, an alternative definition is needed:
\begin{align}
G(\underbrace{0,\dots,0}_{n};x) \,\define\, \frac{\log^n(x)}{n!} \,.
\label{eq:gpldefzeros}
\end{align}
The number of indices $n$ is known as the \emph{weight} of the GPL.
Whenever one of the $a_i$ is on the straight line in the complex plane connecting $0$ and $x$, the integral of eq. \eqref{eq:gpldef} is badly defined and a regularization prescription is needed. Following refs. \cite{Vollinga:2004sn, Frellesvig:2016ske}, we choose
\begin{align}
G(a_1, \ldots, a_n; x) \rightarrow G(a_1, \ldots, a_n; x(1 - i \epsilon))
\label{eq:regularization}
\end{align}
where $\epsilon$ is a small positive number.

A related function is the Multiple Polylogarithm (MPL), which is defined\footnote{We note that unfortunately, the opposite definition is also prevalent in the literature. In that case $\widehat{\Li}_{m_1,\ldots,m_n}(z_1,\ldots,z_n) = \!\! \sum_{0<k_1<\cdots<k_n}^{\infty} \frac{z_1^{k_1}}{k_1^{m_1}} \cdots \frac{z_n^{k_n}}{k_n^{m_n}}$. The relation between the two definitions is $\widehat{\Li}_{m_1,\ldots,m_n}(z_1,\ldots,z_n) = \Li_{m_n,\ldots,m_1}(z_n,\ldots,z_1)$. The ambiguity is inherited by the multiple zeta values, with the alternative definition being $\widehat{\zeta}_{m_1,\ldots,m_n} = \!\! \sum_{0<k_1<\cdots<k_n}^{\infty} \frac{1}{k_1^{m_1} \cdots k_n^{m_n}}$ related to the $\zeta$ of eq. \eqref{eq:mzvdef} as $\widehat{\zeta}_{m_1,\ldots,m_n} = \zeta_{m_n,\ldots,m_1}$} as the sum
\begin{align}
\Li_{m_1,\ldots,m_n}(z_1,\ldots,z_n) &\define \!\! \sum_{k_1>\cdots>k_n>0}^{\infty} \frac{z_1^{k_1}}{k_1^{m_1}} \cdots \frac{z_n^{k_n}}{k_n^{m_n}} \,,
\label{eq:mpldef}
\end{align}
where all $m_i$ are taken to be positive integers. The sum converges whenever $|z_1| \leq 1$, $|z_1 z_2| \leq 1$, $\ldots$, and $|z_1 \cdots z_n| \leq 1$. The number of indices $n$ is known as the \emph{depth} of the MPL, and we note that eq. \eqref{eq:mpldef} is consistent with the definition of the classical polylogarithm $\Li_m(z)$, which the MPL reduces to when the depth is $1$.

The relation between MPLs and GPLs is
\begin{align}
& \Li_{m_1,\ldots,m_n}(z_1,\ldots,z_n) \; =  \nn \\
& (-1)^n G \Big( \underbrace{0,\dots,0}_{m_1-1},\tfrac{1}{z_1},\underbrace{0,\dots,0}_{m_2-1},\tfrac{1}{z_1 z_2},\ldots,\underbrace{0,\dots,0}_{m_n-1},\tfrac{1}{\prod_{i=1}^n \! z_i};1 \Big) .
\label{eq:mplrelation}
\end{align}

Finally the Multiple Zeta Values (MZVs) are defined as
\begin{align}
\zeta_{m_1,\ldots,m_n} &\define \!\! \sum_{k_1>\cdots>k_n>0}^{\infty} \frac{1}{k_1^{m_1} \cdots k_n^{m_n}} \; = \; \Li_{m_1,\ldots,m_n}\big(\underbrace{1,\dots,1}_{n} \big) ,
\label{eq:mzvdef}
\end{align}
with the $m_i$ being positive integers. This definition is consistent with that of the Riemann zeta function evaluated at positive integers $\zeta_m = \zeta(m)$.
The MZVs can be expressed as GPLs as
\begin{align}
\zeta_{m_1,\ldots,m_n} = (-1)^n G \Big( \underbrace{0,\dots,0}_{m_1-1}, 1, \underbrace{0,\dots,0}_{m_2-1}, 1, \ldots,\underbrace{0,\dots,0}_{m_n-1}, 1 ;1 \Big) ,
\label{eq:zetaasGPL}
\end{align}
in accordance with eq. \eqref{eq:mplrelation}.

In addition to the GPL, the MPL, and the MZVs, we will define two additional polylogarithmic functions, both of which are intermediate cases between the GPL/MPL and the classical polylogarithm $\Li_n$. The first is the harmonic polylogarithm\cite{Remiddi:1999ew}
\begin{align}
H(a_1,\ldots,a_n;x) \define (-1)^{\mu} G(a_1,\ldots,a_n;x)\,
\label{eq:hpldef}
\end{align}
where the $a_i$ are taken from the set $\{-1,0,1\}$ and where $\mu$ denotes the number of $a_i$ that equal $1$.
The other such intermediate function is Nielsen's polylogarithm defined as
\begin{align}
S(n,p,z) \define (-1)^{p} G(\underbrace{0,\dots,0}_{n},\underbrace{1,\dots,1}_{p},z)\,.
\label{eq:nielsendef}
\end{align}

A comment on the restriction on the indices of the MPL and the MZV as defined in eqs. \eqref{eq:mpldef} and \eqref{eq:mzvdef} to positive integers is in order: The classical polylogarithm $\Li_n(z)$ and the Riemann zeta function $\zeta(x)$ (as well as Nielsen's polylogarithm mentioned above) are defined for general complex values of all indices and arguments, suggesting that such an extension could be made for the generalized cases as well. Inside the convergent region of the MPL, the extension of the allowed values of the $m_i$ from positive integers to arbitrary complex numbers seems obvious. The reason such cases are not discussed here, nor implemented in Maple, is that the exact nature of the analytical continuation that would be required for such cases, still is an open question mathematically\cite{Goncharov:2001iea}. One reason is that non-integer values of the indices do not allow for the relation to the GPL integral form through eq. \eqref{eq:mplrelation}, and therefore mappings of na{\"i}vely divergent MPLs to the convergent region using integral relations as will be described in section \ref{sec:numerics}, is no longer possible. One additional reason for the restriction to positive integers, is that these cases are the only ones appearing in physics (to the best of the author's knowledge). Generalizations to regions other than positive integers have been discussed in the literature, see e.g. ref. \cite{Goncharov:2001iea}.

\subsection{Divergences}
\label{sec:divergences}

The generalized polylogarithm $G(a_1,\ldots,a_n;x)$ diverges whenever $x=a_1$. The only exceptions to this are $G(1,0,\dots,0;1)$ which evaluates to finite constants, and $G(0,a_2,\ldots,a_n;0)$ which vanishes unless all the $a_i$ equal zero, in which case it does diverge.

The divergence of the GPL is inherited by the MPLs and the MZVs, as \\
$\Li_{1,m_2,\ldots,m_n}(1,z_2,\ldots,z_n)$ is divergent and so is $\zeta_{1,m_2,\ldots,m_n}$.

\subsection{Relations}
\label{sec:relations}

Before listing some of the many relations obeyed by the polylogarithmic functions, we will mention a property that they all share: conservation of weight. For GPLs, the weight is defined as the number of $a_i$ indices, and for the MPLs and the MZVs that corresponds to the sum of the $m$-indices. The definition can be extended such that the product of two objects with weights $w_1$ and $w_2$, will have weight $w_1+w_2$, and in addition rational numbers gets assigned weight $0$. With these definition, all relations listed in this paper conserve this quantity\cite{Goncharov:1998kja, Goncharov:2001iea} in the sense that if the object on the left hand side of an equation has weight $w$, so will each individual term on the right hand side\footnote{Care has to be taken, when integrations or differential operators are involved, or when a relation contains an infinite number of terms.}.

The most significant relation for GPLs, is the rescaling relation
\begin{align}
G(a_1, \ldots, a_n; x) = G(z a_1, \ldots, z a_n; z x)
\label{eq:rescaling}
\end{align}
where $z$ can be any non-zero complex number. The relation is only valid if $a_n \neq 0$.

Additionally there is the shuffle relation\cite{Borwein:1999aa} for the product of two GPLs:
\begin{align}
G(a_1,\ldots,a_m;x) G(b_1, \ldots, b_n; x) &= \sum_{c \in a \sha b} G(c_1, \ldots, c_{m+n}; x)
\label{eq:shuffle}
\end{align}
where $a \sha b$ denotes the shuffles of the lists $a=\{a_1,\ldots,a_m\}$ and $b=\{b_1,\ldots,b_n\}$, which is defined as the set of those of all possible lists containing the elements of $a$ and $b$, for which the ordering of the elements of $a$ and $b$ are the same as in the original lists. The MPLs and the MZVs, inherits the shuffle relation through eqs. \eqref{eq:mplrelation} and \eqref{eq:mzvdef}.

A similar kind of relation, but this time naturally expressed in terms of MPLs, is the stuffle (or quasi-shuffle) identity \cite{Borwein:1999aa, Duhr:2014woa}.
Here the stuffle $\shcha$ of two list is defined as
\begin{align}
a \! \stuffle{\circ} \! b \, &= \, \bigcup_{j=0} \mathcal{M}_{\circ}^j \big( a \sha b \big)
\label{eq:stuffleopdef}
\end{align}
where $\mathcal{M}_{\circ}(x)$ is an operator acting on a list $x$, which returns the set of all lists which may be obtained by taking two adjacent elements of $x$ and replacing them with one element that equals the original two joined by the operator $\circ$, under the condition that one of the two elements come from the list $a$ and the other from $b$.
Given this definition, the stuffle product rule may be expressed as
\begin{align}
\Li_{m_1,\ldots,m_a}(x_1,\ldots,x_a) \Li_{n_1,\ldots,n_b}(y_1,\ldots,y_b) = \sum_{i} \Li_{u_{i}} (z_{i}) \,.
\label{eq:stufflepr}
\end{align}
where $i$ runs over the members of $u = m \!\! \stuffle{+} \!\! n$ and $z = x \!\! \stuffle{\times} \!\! y$. Note that it is important that the lists $u$ and $z$ are sorted in the same way.

The GPLs and MZVs inherits the stuffle rule, through their relations to the MPL.

Another relation is the H{\"o}lder relation (or H{\"o}lder convolution) \cite{Borwein:1999aa, Vollinga:2004sn}
\begin{align}
G(a_1,\ldots,a_n;1) &= \sum_{j=0}^n (-1)^j G(1-a_j,1-a_{j-1},\ldots,1-a_1;1-q) G(a_{j+1},\ldots,a_n;q)
\label{eq:holder}
\end{align}
where $q$ is a number that may take values from a subset of $\mathbb{C}$ that includes the real numbers. The H{\"o}lder relation holds only if $a_1 \neq 1$ and $a_n \neq 0$.

Picking $q=0$ in eq. \eqref{eq:holder}, gives a particularly simple form
\begin{align}
G(a_1,\ldots,a_n;1) &= (-1)^n G(1-a_n,1-a_{n-1},\ldots,1-a_1;1)
\label{eq:holder2}
\end{align}
For the cases of MPLs and MZVs, this is known as the duality relation\cite{Blumlein:2009cf}.

Mapping between MPLs and GPLs is almost one-to-one and is done using eq. \eqref{eq:mplrelation}. The exception is GPLs for which $a_n=0$, which is seen to not correspond to a MPL directly. Such GPLs are said to have \emph{trailing zeros}. If mapping to MPLs is desired for such a case, the zero(s) first have to be removed using the shuffle relation of eq. \eqref{eq:shuffle}. If $a_{n-1} \neq 0$ the relation\cite{Maitre:2005uu} is
\begin{align}
G(a_1,\ldots,a_{n-1},0;x) &= G(a_1,\ldots,a_{n-1};x) G(0;x) - G(0,a_1,\ldots,a_{n-1};x) \nn \\
& \;\;\;\;\; - \ldots - G(a_1,\ldots,a_{n-2},0,a_{n-1};x) \label{eq:trailingzeros}
\end{align}
where each of the terms on the right hand side can be mapped to MPLs, with the exception of $G(0;x)=\log(x)$. For GPLs with more than one trailing zero, a similar procedure may be used.

Another, similar, use of the shuffle product, is for isolating the divergences of a GPL. As described in section \ref{sec:divergences}, a GPL $G(x, a_2, \ldots, a_n; x)$ will in general diverge. The divergent part can, however, always be isolated as powers of a divergent logarithm $G(x; x) = \log(0)$. This can be done using the shuffle rules as above, an example (valid for $a_2 \neq x$) being
\begin{align}
G(x,a_2,\ldots,a_{n};x) &= G(a_2,\ldots,a_{n};x) G(x;x) - G(a_2,x,a_3,\ldots,a_{n};x) \nn \\
& \;\;\;\;\; - \ldots - G(a_2,\ldots,a_{n},x;x). \label{eq:isolatedivergences}
\end{align}
A similar procedure can be applied when the list of indices of the GPL, begins with more than one $x$.

It should be mentioned that generalized polylogarithms have a large number of relations between each other in addition to those listed in this section (which are those implemented in Maple). See e.g. refs. \cite{lewin1981, Goncharov:1998kja, Goncharov:2010jf, Borwein:1999aa, Vollinga:2004sn, Duhr:2012po, Ablinger:2013cf, Panzer:2014caa, Frellesvig:2016ske}.

\subsection{Special values}
\label{sec:specialvalues}

At certain special values of the arguments, GPLs, MPLs, and MZVs can be expressed in terms of simpler functions.

Whenever the last argument of a GPL is zero, it vanishes\footnote{with the exceptions of $G(;0) = 1$, and $G(0,\ldots,0;0)$ which diverges.}, and so does MPLs with any zero argument:
\begin{align}
G(a_1,\ldots,a_n;0) = 0 \,, \quad\quad\;\; \Li_{m_1,\ldots,m_n}(z_1,\ldots,z_{k-1},0,z_{k+1},\ldots,z_n) = 0 \,.
\label{eq:evaluatestozero}
\end{align}

For GPLs a simple relation is given by eq. \eqref{eq:gpldefzeros}, and additional simple relations are
\begin{align}
G(\underbrace{a,\dots,a}_{n};x) &= \frac{\log^n \! \left( 1- \tfrac{x}{a} \right) }{n!} \label{eq:gsimp1} \\
G(\underbrace{0,\dots,0}_{n-1},a;x) &= - \Li_n \! \left( \tfrac{x}{a} \right) \label{eq:gsimp2}
\end{align}
For the case of MPLs, eq. \eqref{eq:gsimp2} becomes the relation to the classical polylogarithm, while eq. \eqref{eq:gsimp1} has no such simple interpretation.

MPLs for which all arguments equal one, are given by MZVs as by eq. \eqref{eq:mzvdef}. If instead the arguments are taken from the set $\{1,-1\}$ the values are known as \emph{oscillating multiple zeta values}\cite{Blumlein:2009cf}.
Examples of oscillating multiple zeta values, not covered by the above relations, are
\begin{align}
& \Li_{1,1,1}(-1,1,-1) = \frac{1}{8} \zeta_3 - \frac{1}{6} \log^3(2), \label{eq:omzv1} \\
& \Li_{2,2}(1,-1) = 4 \Li_4 \! \big( \tfrac{1}{2} \big) + \frac{1}{6} \log^4(2) - \frac{\pi^2}{6} \log^2(2) + \frac{7}{2} \zeta_3 \log(2) - \frac{71}{1440} \pi^4. \label{eq:omzv2}
\end{align}
The GPLs inherits these values from the MPLs through eq. \eqref{eq:mplrelation}.
A GPL without trailing zeros, for which all $a_i$ equal either $0$ or $x$ will be given by a MZV, while a GPL without trailing zeros for which all $a_i \in \{0,x,-x\}$ will be given by an oscillating multiple zeta value.

In general the shuffle and stuffle relations together allows for the mapping of GPLs to a minimal set of independent functions\cite{Duhr:2012po}. For weight $\leq 4$, that set has been shown\cite{Frellesvig:2016ske} to consist of the functions $\Li_n(x)$ and $\Li_{2,2}(x,y)$. In general such mappings will contain a number of step-functions in order to pick the right branches of the polylogarithmic functions, but when the arguments are taken from the set of zero along with at most two constants, one expression can be shown to cover all of argument space. At weight two the non-trivial (in the sense of not being covered by previous relations) identities are
\begin{align}
G(a,0;x) &= \Li_2 \! \left( \tfrac{x}{a} \right) + \log(x) \log \! \left( 1 - \tfrac{x}{a} \right), \label{eq:gplspecial1} \\
G(a,x;x) &= -\Li_2 \! \left( \tfrac{x}{x-a} \right), \label{eq:gplspecial2}
\end{align}
while at weight three, an example is
\begin{align}
G(a,x,a;x) &= -2 \, \Li_3 \! \left( \tfrac{a}{a-x} \right) - \Li_2 \! \left( \tfrac{a}{a-x} \right) \log \! \left( 1 - \tfrac{x}{a} \right) - \frac{\pi^2}{6}  \log \! \left( 1 - \tfrac{x}{a} \right) + 2 \zeta_3.  \label{eq:gplspecial3}
\end{align}

For MZVs, there are multiple additional relations that would qualify as special values, some of which are without direct analogues for MPLs or GPLs. Examples are
\begin{align}
\zeta_{m,m} &= \big( \zeta_m^2 - \zeta_{2m} \big)/2,  \label{eq:zetasimp1} \\
\zeta_{2,1,\ldots,1} &= \zeta_{d+1}, \label{eq:zetasimp2}
\end{align}
where $d$ denotes the depth (i.e. the number of arguments) of the $\zeta$ on the left hand side. Eq. \eqref{eq:zetasimp2} is a special case of the duality relation mentioned above as eq. \eqref{eq:holder2}.

Another relation\cite{PanzerKosmos}, known as a parity relation, is
\begin{align}
\zeta_{m,n} = (-1)^m \!\!\!\! \sum_{s=0}^{(\nu-3)/2} \!\! \left( \bin{\nu-2s-1}{m-1} + \bin{\nu-2s-1}{n-1} - \delta_{2s,n} + (-1)^m \delta_{s,0} \right) \zeta_{2s} \zeta_{\nu-2s}
\label{eq:oddzetarelation}
\end{align}
where $\nu = n+m$ has to be odd. To use eq. \eqref{eq:oddzetarelation} we need to use $\zeta_0 \define \zeta(0) = -\tfrac{1}{2}$.

As for the case of GPLs, the shuffle and stuffle relations together allow for the reduction of all MZVs up to a given weight, to linear combinations of a small subset thereof. For weights $\leq 7$ that set consists of only the standard zeta values, e.g. the Riemann zeta function evaluated at positive integers. As even zeta values (i.e. $\zeta_{2n}$) are proportional to powers of $\pi$ with rational coefficients and therefore to each other:
\begin{align}
\zeta_2 = \frac{\pi^2}{6} \;, \quad\quad \zeta_4 = \frac{\pi^4}{90} = \frac{2}{5} \zeta_2^2 \;,\quad\quad \zeta_6 = \frac{\pi^6}{945} = \frac{8}{35} \zeta_2^3 \;, \ldots
\end{align}
this means that all multiple zeta values for weight $\leq 7$ can be expressed in terms of only the four zeta values $\pi^2/6$, $\zeta_3$, $\zeta_5$, $\zeta_7$.
Examples are
\begin{align}
\zeta_{3,1,2} &= \frac{53}{22680} \pi^6 - \frac{3}{2} \zeta_3^2 \,,  \label{eq:zetared1} \\
\zeta_{4,2,1} &= \frac{7}{180} \pi^4 \zeta_3 + \frac{11}{12} \pi^2 \zeta_5 - \frac{221}{16} \zeta_7 \,.  \label{eq:zetared2}
\end{align}
At higher weights that is no longer the case, at weight $8$ one double zeta value has to be added to the basis, and if $\zeta_{5,3}$ is chosen, an example is
\begin{align}
\zeta_{2,3,3} &= \frac{793}{1134000} \pi^8 + \frac{1}{3} \pi^2 \zeta_3^2 - 9 \zeta_3 \zeta_5 - \frac{27}{10} \zeta_{5,3} \,,  \label{eq:zetared3}
\end{align}
and further double and multiple zeta values have to be added to the basis at higher weights.

\subsection{Derivatives}
\label{sec:derivatives}

The derivative of a GPL with respect to the argument follows trivially from eq. \eqref{eq:gpldef} and is given by
\begin{align}
\frac{\partial}{\partial x} G(a_1, \ldots, a_n; x) &= \frac{G(a_2, \ldots, a_n; x)}{x - a_1}
\label{eq:gdiffx}
\end{align}
both for arbitrary and zero-valued $a_i$.

A derivative with respect to one of the indices is more intricate\cite{Vollinga:2004sn}. When there is only one index, $G(a;x) = \log(1-x/a)$, and the result is
\begin{align}
& \frac{\partial}{\partial a} G(a; x) = \frac{-1}{x-a} - \frac{1}{a} \,.
\label{eq:gdiff1}
\end{align}
For a general GPL, the derivative with respect to the first index is
\begin{align}
& \frac{\partial}{\partial a_1} G(a_1, a_2, a_3, \ldots, a_n; x) \label{eq:gdiff2} \\
&= \left( \frac{1}{a_2-a_1} - \frac{1}{x - a_1} \right) G(a_2,a_3, \ldots, a_n; x) - \frac{1}{a_2-a_1} G(a_1,a_3, \ldots, a_n; x), \nn
\end{align}
the derivative with respect to the last index is
\begin{align}
& \frac{\partial}{\partial a_n} G(a_1, \ldots, a_{n-2}, a_{n-1}, a_n; x) \label{eq:gdiff3} \\
&= \left( \frac{1}{a_{n}-a_{n-1}} - \frac{1}{a_n} \right) G(a_1, \ldots, a_{n-2}, a_{n-1}; x) - \frac{1}{a_n-a_{n-1}} G(a_1, \ldots, a_{n-2}, a_n; x), \nn
\end{align}
and for an intermediate index, we get
\begin{align}
& \frac{\partial}{\partial a_k} G(a_1, \ldots, a_{k-1}, a_{k}, a_{k+1}, \ldots, a_n; x) \nn \\
&= \left( \frac{1}{a_{k+1}-a_{k}} + \frac{1}{a_k - a_{k-1}} \right) G(a_1, \ldots, a_{k-1}, a_{k+1}, \ldots, a_n; x) \label{eq:gdiff4} \\
& - \frac{1}{a_{k+1}-a_{k}} G(a_1, \ldots, a_{k}, a_{k+2}, \ldots, a_n; x) - \frac{1}{a_{k}-a_{k-1}} G(a_1, \ldots, a_{k-2}, a_{k}, \ldots, a_n; x). \nn
\end{align}

It may appear from eqs. \eqref{eq:gdiff2} to \eqref{eq:gdiff4}, that singularities will appear when two adjacent indices are identical. But that merely reflect the fact that a partial derivative does not capture the full functional dependence in such cases. In that case one needs to take the derivative with respect to both of the identical indices, and then the apparent singularity will cancel: 
\begin{align}
& \frac{d}{d y} G(a_1, \ldots, a_{k-1}, y, y, a_{k+2}, \ldots, a_n; x) \nn \\
&= \left( \left( \frac{\partial}{\partial a_k} + \frac{\partial}{\partial a_{k+1}} \right) G(a_1, \ldots, a_{k-1}, a_k, a_{k+1}, a_{k+2}, \ldots, a_n; x) \right) \bigg|_{a_{k} = a_{k+1} = y} \nn \\
&= \left( \frac{1}{y - a_{k-1}} + \frac{1}{a_{k+2}-y} \right) G(a_1,\ldots,a_{k-1},y,a_{k+2},\ldots,a_n;x) \label{eq:gdiff5} \\
& \;\;\;\;\;\; - \frac{1}{a_{k+2}-y} G(a_1, \ldots, a_{k-1}, y, y, a_{k+3}, \ldots, a_n; x) \nn \\
& \;\;\;\;\;\; - \frac{1}{y-a_{k-1}} G(a_1, \ldots, a_{k-2}, y, y, a_{k+2}, \ldots, a_n; x). \nn
\end{align}

Derivatives of MPLs with respect to the arguments, should be done by mapping them to GPLs using eq. \eqref{eq:mplrelation}, do the derivative there using the above relations, and then map back.

Derivatives of MPLs or MZVs with respect to the indices are not defined, as these indices only take values in the positive integers and therefore the functional dependence thereon is not differentiable.

\subsection{Symbol algebra}
\label{sec:symbols}

For applications in physics, it is common to perform manipulations and simplifications of expressions containing GPLs and related functions, using an algebraic object known (in physics) as the \emph{symbol} \cite{Goncharov:2010jf, Duhr:2012po}. Symbol is short for Chen symbol after ref. \cite{chen1977}. The symbol is designed to capture the algebraic parts of relations between polylogarithmic functions, while ignoring the analytic parts of such relations, such as branch cuts.

For a function $f(x)$ of a set of variables $x$, for which the total derivative may be expressed as
\begin{align}
\id f(x) = g(x) \, \id \! \log(h(x)), 
\label{eq:symdef1}
\end{align}
the symbol $\mathcal{S}$ of the function is defined recursively as
\begin{align}
\mathcal{S}(f(x)) = \mathcal{S}(g(x)) \otimes h(x).
\label{eq:symdef2}
\end{align}
The recursion stops as $\mathcal{S}(\log(x)) = x$.

For the GPL, this gives the symbol
\begin{align}
\mathcal{S}(G(a_1,\ldots,a_n;x)) &= \sum_{i=1}^{n} \bigg( \mathcal{S}\big( G(a_1,\ldots,\hat{a}_i,\ldots,a_n;x) \big) \otimes (a_i - a_{i-1}) \nn \\
& \quad\quad - \mathcal{S}\big( G(a_1,\ldots,\hat{a}_i,\ldots,a_n;x) \big) \otimes (a_i - a_{i+1}) \bigg),
\end{align}
where $a_{n+1} \define 0$ and $a_{0} \define x$, and where $\hat{a}_i$ indicates that the $a_i$ entry is left out.

The symbol operator $\mathcal{S}$ obeys
\begin{align}
\mathcal{S}(q f(x)) &= q \mathcal{S}(f(x)) \label{eq:symrules1} \\
\mathcal{S}(f(x) + g(x)) &= \mathcal{S}(f(x)) + \mathcal{S}(g(x)) \label{eq:symrules2} \\
\mathcal{S}(f(x) g(x)) &= \mathcal{S}(f(x)) \sha \mathcal{S}(g(x)) \label{eq:symrules3}
\end{align}
where $q$ is a rational number, and where $\sha$ denotes the shuffle operator defined in section \ref{sec:relations}.

The symbol of certain transcendental constants, such as MZVs vanish:
\begin{align}
\mathcal{S}(\zeta_{m_1,\ldots,m_n}) = 0 \,, \quad\quad\;\; \mathcal{S}(i \pi) = 0,
\end{align}
but please note that not all symbols of constants have to vanish, it is fully consistent\cite{Duhr:2012po} to put for instance
\begin{align}
\mathcal{S}(\log(q)) = q \quad\quad\;\; (q \in \mathbb{Q}).
\label{eq:numberinsymbol}
\end{align}
Other simple special cases are
\begin{align}
\mathcal{S}(\Li_n(x)) &= - \Big( (1-x) \otimes \underbrace{x \otimes \cdots \otimes x}_{n-1} \Big), \\
\mathcal{S}(\log^n(x)) &= n! \Big( \underbrace{x \otimes \cdots \otimes x}_{n} \Big).
\end{align}

The symbol itself, also obeys various relations, that it inherits from the logarithmic differential of eq. \eqref{eq:symdef1}:
\begin{align}
\Big( \cdots \otimes xy \otimes \cdots \Big) &= \Big( \cdots \otimes x \otimes \cdots \Big) + \Big( \cdots \otimes y \otimes \cdots \Big)\,, \\
\Big( \cdots \otimes x^q \otimes \cdots \Big) &= q \Big( \cdots \otimes x \otimes \cdots \Big) \quad\quad (q \in \mathbb{Q})\,,\\
\Big( \cdots \otimes 1 \otimes \cdots \Big) &= 0\,, \\
\Big( \cdots \otimes -x \otimes \cdots \Big) &= \Big( \cdots \otimes x \otimes \cdots \Big)\,.
\end{align}

We will not here do any examples of how the use the symbol to simplify calculations with GPLs. For summaries, see e.g. refs. \cite{Duhr:2012po, Duhr:2012fh, Duhr:2014woa}.

\section{Numerical evaluation}
\label{sec:numerics}

The algorithm for numerical evaluation of generalized polylogarithms discussed in this section, follows largely the algorithm discussed in ref. \cite{Vollinga:2004sn}.

The sum defining for the MPL $\Li_{m_1,\ldots,m_n}(z_1,\ldots,z_n)$, given by eq. \eqref{eq:mpldef}, converges whenever $|z_1| \leq 1$, $|z_1 z_2| \leq 1$, $\ldots$, $|z_1 \cdots z_n| \leq 1$, as described in section \ref{sec:definitions}. For GPLs $G(a_i,\ldots,a_n;x)$, this criterion translates to a requirement that $\forall a_i|_{a_i \neq 0}: |a_i| \geq |x|$. When that is not the case, the GPL in question has to be mapped to GPLs that have this property, for instance using the following algorithm:

Let us consider a GPL for which the argument with the smallest non-zero absolute value is not $x$ but one of the $a_i$, which we will denote $s$. The goal is now to map this GPL to GPLs with arguments taken from the same set, but where $s$ is either at the last position so the GPL converges, or not present at all in which case the argument with the smallest non-zero value will be a different (but larger) one, and the algorithm will have to be repeated. All the cases discussed below are for GPLs without trailing zeros. If trailing zeros are present, they first have to be removed using the algorithm described in and around eq. \eqref{eq:trailingzeros}.

The simplest imaginable case is $G(s,x)$. Using eq. \eqref{eq:gsimp1}, we get
\begin{align}
G(s;x) &= \log \! \left( 1 - \tfrac{x}{s} \right) \nn \\
& = G(x;s) - G(0;s) + G(0,-x) + 2 \pi i \Phi \! \left( -x, \tfrac{1}{s} \right) \label{eq:ssc1}
\end{align}
where $\Phi$ is a function of the complex phases of its arguments, designed to have the property
\begin{align}
\log(ab) &= \log(a) + \log(b) + 2 \pi i \Phi(a,b)
\label{eq:phidef}
\end{align}
for any values of $a$ and $b$. We see that eq. \eqref{eq:ssc1} have the required property that all GPLs on the right hand side are either independent of $s$, or have $s$ as its last argument.

The second-simplest case is GPLs for which $s$ is the last of the $a_i$ arguments, but the remaining $a_i$s equal zero. In that case we may use eq. \eqref{eq:gsimp2} to map to the classical polylogarithm $\Li_n(x/s)$, and then use the inversion relation for $\Li_n$ \cite{lewin1981} \\[-7mm]
\begin{align}
\Li_n(z) = (-1)^{n-1} \Li_n \! \left( \tfrac{1}{z} \right) + 2 \sum_{r=0}^{\floor{\tfrac{n}{2}}} \frac{\log^{n-2r}(-z)}{(n-2r)!} \left( 2^{1-2r} - 1 \right) \zeta(2r)
\label{eq:lininversion}
\end{align}
The $\Li_n$ appearing on the right hand side of eq. \eqref{eq:lininversion} can be expressed as the GPL $G(0,\ldots,0,x;s)$, and the logarithms can be written as
\begin{align}
\log \! \left( - \tfrac{x}{s} \right) = - G(0;s) + G(0,-x) + 2 \pi i \Phi \! \left( -x, \tfrac{1}{s} \right)
\end{align} 
so we see that all terms on the right hand side will have the desired form.

The next case we will consider, is cases where $s$ is the last of the $a_i$, but where not all of the remaining $a_i$ are zero. For a GPL where $s$ is preceded by $\nu-1$ zeros (i.e. zero or more), we will re-express it as 
\begin{align}
G(a_1,\ldots,a_k,\underbrace{0,\dots,0}_{{\nu}-1},s;x) &= G(a_1,\ldots,a_k;x) G(\underbrace{0,\dots,0}_{{\nu}-1},s;x) \label{eq:ssc3} \\[-2mm]
& \;\;\;\;\; - \sum \text{remaining shuffle terms} \nn
\end{align}
where ``shuffle terms'' refers to all the other terms one would obtain from applying the shuffle product rule eq. \eqref{eq:shuffle} to $G(a_i,\ldots,a_k;x) G(0,\dots,0,s;x)$. Those shuffle terms would all either have $s$ as the last $a_i$ but preceded by less than $\nu-1$ zeros in which case eq. \eqref{eq:ssc3} should be applied again recursively, or it would not have $s$ as the last of the $a_i$ in which case the following paragraph applies.

The last case, is the case where $s$ is not the last of the $a_i$ indices. For such cases we utilize the obvious relation
\begin{align}
& G(a_1,\ldots,a_{k-1},s,a_{k+1},\ldots,a_n;x) = \label{eq:ssc4} \\
& G(a_1,\ldots,a_{k-1},0,a_{k+1},\ldots,a_n;x) + \int_0^s \! \frac{\partial}{\partial t} G(a_1,\ldots,a_{k-1},t,a_{k+1},\ldots,a_n;x) \, \id t \nn
\end{align}
Here the first term is independent of $s$, and for the second one, the derivative can be evaluated\cite{Vollinga:2004sn, Panzer:2014caa} using the expressions listed in section \ref{sec:derivatives} such as eqs. \eqref{eq:gdiff2} and \eqref{eq:gdiff4}. Performing the derivative will give some terms for which the $t$-integration has to be done over a term of the form $1/(t-a)$ times a GPL with no dependence at all on $t$, which is easily evaluated to the GPL times another GPL of weight one. There will also be terms where the integration has to be done over the $1/(t-a)$ factor times a GPL that retain the dependence on $t$, but which has a weight that is lowered by one compared to the original. For such terms eq. \eqref{eq:ssc4} should be applied recursively until the integration is over $1/(t-a)$ multiplied either with constants or with GPLs that have $t$ as its last argument, in which case the integration can be done using eq. \eqref{eq:gpldef}.

The procedure for mapping a given argument of a GPL to the last position, described in eqs. \eqref{eq:ssc1} to \eqref{eq:ssc4}, are occasionally referred to as ``super shuffle identities''.

The above procedure depends on the ability to distinguish the phases and the absolute values of different indices of the GPL. But numerically it may well happen that arguments have the same absolute value or the same phase, so to avoid ambiguities, factors corresponding to infinitesimal changes of arguments $(1 - i \epsilon)$ or absolute values $(1 - \epsilon)$ are multiplied on when necessary, in accordance with the regularization prescription given by eq. \eqref{eq:regularization}.

The rate of convergence of the series of eq. \eqref{eq:mpldef} is heavily dependent on the values of the terms in the numerator. In the language of GPLs, convergence requires that $|x| \leq$ the absolute value of the smallest of the non-zero $a_i$. But if the two values are close, the convergence may be very slow and an alternative method is desirable. The MZVs form a limiting case, in the sense that $x$ and all non-zero $a_i$ are identical for that case.

One such method\cite{Vollinga:2004sn} is to rescale $x$ to one, and then apply the H{\"o}lder convolution of eq. \eqref{eq:holder} with $q=\tfrac{1}{2}$. Doing so will map most badly convergent cases to a combination of GPLs that are all either well convergent (for MZVs this includes all terms), or non-convergent in a way that can be mapped to well convergent cases by the algorithm described in the beginning of this section. For a few unlucky cases, applying the H{\"o}lder convolution a second time may be needed \cite{Vollinga:2004sn}.

\section{Implementation in Maple 2018}
\label{sec:implementation}

In this section we will describe the Maple functions {\tt GeneralizedPolylog}, {\tt MultiPolylog}, and {\tt MultiZeta}, as well as the Maple package {\tt PolylogTools} containing tools for the manipulation thereof.

\subsection{GeneralizedPolylog}
\label{sec:implGPL}

{\tt GeneralizedPolylog} is the name of the Maple implementation of the generalized polylogarithm as defined in eq. \eqref{eq:gpldef}.
The function should be called as
\begin{verbatim}
> GeneralizedPolylog(a::list, x)
\end{verbatim}
where {\tt x} and the elements of {\tt a} should be Maple constants.

When called, {\tt GeneralizedPolylog} first checks for singularities. If there is a singularity according to the criterion given in section \ref{sec:divergences}, {\tt GeneralizedPolylog} will return an error. Otherwise, it will continue by looking for special values. Expressions for the special values are described in section \ref{sec:specialvalues}, and they are implemented in the following order:
$ $\\[0mm]

\begin{tabular}{| l l |}
\hline
$\;\;\;$ Special values & $\;\;\;$ reference \\ \hline
$G(;x) = 1$ & $\quad-$ \\
$G(a_1,\ldots,a_n;0) = 0$ & eq. \eqref{eq:evaluatestozero} \\
Logarithms & eqs. \eqref{eq:gpldefzeros} and \eqref{eq:gsimp1} \\
Classical polylogarithms & eq. \eqref{eq:gsimp2} \\
Multiple zeta values (MZVs) & eq. \eqref{eq:zetaasGPL} and section \ref{sec:implMZV} \\
Oscillating multiple zeta values & eqs. \eqref{eq:omzv1} and \eqref{eq:omzv2}, and ref. \cite{Blumlein:2009cf} \\
Special reductions & eqs. \eqref{eq:gplspecial1} to \eqref{eq:gplspecial3}, and ref. \cite{Frellesvig:2016ske} \\ \hline
\end{tabular}
$ $\\[2mm]

The mappings to logarithms, classical polylogarithms, and MZVs are applied whenever possible. The oscillating multiple zeta values are implemented as a table up to weight six, but by default the mappings are only applied for weights $\leq 4$, as they do not provide much simplification at higher weights. The ``special reductions'' are implemented for all applicable cases at weights $\leq 3$.

The numerical evaluation of GPLs is done by the Maple function\\ {\tt EvalfGeneralizedPolylog(a::list, x)}, which applies the algorithm described in section \ref{sec:numerics}. When {\tt evalf} is called on a {\tt GeneralizedPolylog}, {\tt EvalfGeneralizedPolylog} gets called automatically. As usual\\ {\tt evalf(GeneralizedPolylog(a,x))} applies the special values mentioned above before doing the numerical evaluation, so if the user wants to access the numerics directly, they will have to call {\tt EvalfGeneralizedPolylog} specifically.

A number of relations for the GPL, are implemented in the {\tt Transformations} environment used for special functions in Maple\cite{ChebTerrab:Wizard}. They are called as 
\begin{verbatim}
> GeneralizedPolylog:-Transformations["Name"][number]( ... )
\end{verbatim}
Eight such transformations are implemented as listed below.
As the names suggest, some identities are implemented in several ways in order for them to be as useful as possible. \\

{\tt Transformations["Rescaling"][1](a::list, x)} applies the rescaling relation eq. \eqref{eq:rescaling} to $G(a_1,\ldots,a_n;x)$. That relation is only valid when $a_n \neq 0$, otherwise the transformation will return nothing. The $q$ appearing on the right hand side of eq. \eqref{eq:rescaling}, is implemented as a local variable in {\tt GeneralizedPolylog}: {\tt GeneralizedPolylog:-q}.

{\tt Transformations["Shuffle"][1](a::list, x, b::list, y)} applies the shuffle relation eq. \eqref{eq:shuffle} to $G(a_1,\ldots,a_n;x) G(b_1,\ldots,b_m;y)$. The relation is only valid if $x=y$, and otherwise the relation will return nothing.

{\tt Transformations["Shuffle"][2](a::list, x, b::list, y)} also applies the shuffle relation eq. \eqref{eq:shuffle} to $G(a_1,\ldots,a_n;x) G(b_1,\ldots,b_m;y)$, but is a more flexible and general implementation. If none of the GPLs have trailing zeros, the rescaling identity is applied, rescaling both GPLs to GPLs with argument $q$, where the $q$ local to {\tt GeneralizedPolylog} mentioned above is used, and then the shuffle product rule is applied. If just one of the GPLs have trailing zeros, the other GPL will get rescaled such that its argument is the same as the argument of the GPL with the trailing zeros, and then the shuffle rule is applied. If both GPLs have trailing zeros, the behaviour is the same as for {\tt Transformations["Shuffle"][1]}.

{\tt Transformations["Shuffle"][3](a::list, x)} uses the shuffle identity to re-express $G(a_1,\ldots,a_n;x)$ in terms of $G(a_1,x)$ and GPLs where the first index is different from $a_1$. This is similar to the algorithm for the isolation of divergences described in and around eq. \eqref{eq:isolatedivergences}, except that no divergences have to be present.

{\tt Transformations["Shuffle"][4](a::list, x)} uses the shuffle identity to re-express $G(a_1,\ldots,a_n;x)$ in terms of $G(a_n,x)$ and GPLs where the last index is different from $a_n$. This is similar to the algorithm for the removal of trailing zeros described in and around eq. \eqref{eq:trailingzeros}, except that $a_n$ does not have to equal zero.

{\tt Transformations["Stuffle"][1](a::list, x, b::list, y)} applies the stuffle product rule given by eq. \eqref{eq:stufflepr} to $G(a_1,\ldots,a_n;x) G(b_1,\ldots,b_m;y)$. This is done by first mapping the two GPLs to MPLs using eq. \eqref{eq:mplrelation}, applying the stuffle product rule, and then mapping the resulting MPLs back to GPLs. As the mapping to MPLs only works directly when no trailing zeros are present, nothing will be returned otherwise.

{\tt Transformations["Holder"][1](a::list, x)} applies the H{\"o}lder identity eq. \eqref{eq:holder} to $G(a_1,\ldots,a_n;x)$. Before the H{\"o}lder identity is applied, the rescaling identity is used to give the GPL argument $1$. For the $q$ appearing on the right hand side of eq. \eqref{eq:holder}, the local variable {\tt GeneralizedPolylog:-q} is used. As eq. \eqref{eq:holder} is valid only when $a_1 \neq x$ and $a_n \neq 0$, nothing will be returned otherwise.

{\tt Transformations["Holder"][2](a::list, x)} applies the H{\"o}lder identity for the special case with $q=0$, as it is given in eq. \eqref{eq:holder2}. As that identity is valid only when $a_1 \neq x$ and $a_n \neq 0$, nothing will be returned otherwise.

An interface to the transformations listed above, is provided by the Maple {\tt Identities} environment\cite{ChebTerrab:Wizard}. It may be called as either
\begin{verbatim}
> GeneralizedPolylog:-Identities(a::list, x)
\end{verbatim}
or
\begin{verbatim}
> GeneralizedPolylog:-Identities(a::list, x, b::list, y)
\end{verbatim}
i.e. with either two or four arguments. This function lists those of the transformations listed above, that are valid for the given arguments, along with additional requirements on their validity.

\subsection{MultiPolylog}
\label{sec:implMPL}

{\tt MultiPolylog} is the Maple implementation of the multiple polylogarithm as defined by eq. \eqref{eq:mpldef}.
The function should be called as
\begin{verbatim}
> MultiPolylog(m::list, z::list)
\end{verbatim}
where {\tt m} and {\tt z} should have the same length, and where the elements of {\tt m} should be consistent with being positive integers.

When called, {\tt MultiPolylog} first checks if its two arguments are lists of the same length, and if the first has members consistent with being positive integers. If not an error is returned. Then it checks for singularities. If there is a singularity according to the criterion given in section \ref{sec:divergences} (i.e. if {\tt m[1]} $ = 1$ and {\tt z[1]} $ = 1$), {\tt MultiPolylog} will also return an error. Otherwise, it will search for special values. Expressions for the special values are described in section \ref{sec:specialvalues}, and they are implemented in the following order:
$ $\\[0mm]

\begin{tabular}{| l l |}
\hline
$\;\;\;$ Special values & $\;\;\;$ reference \\ \hline
$\Li() = 1 \;\;$ (no arguments) & $\quad-$ \\
$\Li_{m_1,\ldots,m_n}(z_1,\ldots,0,\ldots,z_n)=0$ & eq. \eqref{eq:evaluatestozero} \\
Classical polylogarithms & $\quad-$ \\
Multiple zeta values (MZVs) & eq. \eqref{eq:mzvdef} and section \ref{sec:implMZV} \\
Logarithms & eq. \eqref{eq:gsimp1} \\
Oscillating multiple zeta values $\;\;\;$ & eqs. \eqref{eq:omzv1} and \eqref{eq:omzv2}, and ref. \cite{Blumlein:2009cf} \\
Special reductions & eqs. \eqref{eq:gplspecial1} to \eqref{eq:gplspecial3}, and ref. \cite{Frellesvig:2016ske} \\ \hline
\end{tabular}
$ $\\[2mm]

As for the GPL implementation described above, the cases of classical polylogarithms, logarithms, and multiple zeta values, are used whenever applicable. The oscillating multiple zeta values are tabulated up to weight six, but are only by default applied for weights $\leq 4$. The special reductions are applied whenever applicable, at weights $\leq 3$.

The numerical evaluation of MPLs are implemented as the procedure\\ {\tt EvalfMultiPolylog(m::list, z::list)}, which is a member of {\tt EvalfGeneralizedPolylog} as it is the same evaluation procedure that is applied, but which can be called as if it were an ordinary function. Applying {\tt evalf} to {\tt MultiPolylog} will call {\tt EvalfMultiPolylog}. As for other cases, {\tt evalf(MultiPolylog(m,z))} will first apply the above mappings to special values before performing the numerics.

As for the GPLs, a number of transformations valid for MPLs are implemented using the Maple {\tt Transformations} environment. These are:

{\tt Transformations["Stuffle"][1](m::list, z::list, u::list, y::list)} applies the stuffle product identity eq. \eqref{eq:stufflepr} to $\Li_{m_1,\ldots,m_n}(z_1,\ldots,z_n) \Li_{u_1,\ldots,u_{\nu}}(y_1,\ldots,y_{\nu})$. The only requirement is that the lengths of the lists are pairwise identical, as required by MPLs.

{\tt Transformations["Shuffle"][1](m::list, z::list, u::list, y::list)} applies the shuffle product identity eq. \eqref{eq:shuffle} to $\Li_{m_1,\ldots,m_n}(z_1,\ldots,z_n) \Li_{u_1,\ldots,u_{\nu}}(y_1,\ldots,y_{\nu})$, by mapping to GPLs, applying the identity, and mapping back. Besides from the requirement that the lengths of the lists are pairwise identical, the only requirement is that both the MPLs are non-divergent.

{\tt Transformations["Duality"][1](m::list, z::list)} applies the duality relation eq. \eqref{eq:holder2} to $\Li_{m_1,\ldots,m_n}(z_1,\ldots,z_n)$. Beside from the usual requirement that the lists have the same length, the only requirement for the duality to be valid, is that the MPL is non-divergent, and the the opposite case nothing is returned.

The transformations can be accessed using the {\tt Identities} environment as described in the previous section.

\subsection{MultiZeta}
\label{sec:implMZV}

{\tt MultiZeta} is the Maple implementation of the multiple zeta values defined by eq. \eqref{eq:mzvdef}.
The function should be called as
\begin{verbatim}
> MultiZeta(m1,...,mn)
\end{verbatim}
i.e. with a variable number of arguments from zero and up, each of which should be consistent with being a positive integer. 

When called, {\tt MultiZeta} first checks if each argument is consistent with being a positive integer. If not an error is returned. Then it checks for singularities. If there is a singularity according to the criterion given in section \ref{sec:divergences} (i.e. if {\tt m1} $= 1$), {\tt MultiZeta} will also return an error. Otherwise, it will continue by looking for special values. Expressions for the special values are described in section \ref{sec:specialvalues}, and they are implemented in the following order:
$ $\\[0mm]

\begin{tabular}{| l l |}
\hline
$\;\;\;$ Special values & $\;\;\;$ reference \\ \hline
$\zeta = 1 \;\;$ (no arguments) & $\quad-$ \\
Depth $1$ $\rightarrow \zeta(m)$ & $\quad-$ \\
$\zeta_{m,m}$ & eq. \eqref{eq:zetasimp1} \\
Depth $w-1$, duality & eq. \eqref{eq:zetasimp2} \\
$\zeta_{m,n}$, $w$ odd & eq. \eqref{eq:oddzetarelation} \\
Tabulated reductions $\;\;\;\;\;$ & eqs. \eqref{eq:zetared1} to \eqref{eq:zetared3}, and ref. \cite{Blumlein:2009cf} \\ \hline
\end{tabular}
$ $\\[2mm]

The ``tabulated reductions'' are tabulated up to weight $10$, but are only implemented by default for weights $ \leq 7$, as higher weights will not in general reduce to inherently simpler functions (see e.g. eq. \eqref{eq:zetared3}). The remaining relations are applied in all applicable cases.

The numerical evaluation is implemented as the function \\ {\tt EvalfMultiZeta(m1,...,mn)}, which works using the algorithm described in section \ref{sec:numerics}. As for the GPL and MPL mentioned above, calling {\tt evalf} on {\tt MultiZeta} will call {\tt EvalfMultiZeta} but only after applying the special values listed in the above table.

As for the GPLs and MPLs, a number of transformations valid for MZVs are implemented using the Maple {\tt Transformations} environment. These are the same as for the MPL case, and are as follows:

{\tt Transformations["Stuffle"][1](m::list,u::list)} applies the stuffle product identity eq. \eqref{eq:stufflepr} to $\zeta_{m_1,\ldots,m_n} \zeta_{u_1,\ldots,u_{\nu}}$. This transformation is valid in all cases.

{\tt Transformations["Shuffle"][1](m::list, u::list)} applies the shuffle product identity eq. \eqref{eq:shuffle} to  $\zeta_{m_1,\ldots,m_n} \zeta_{u_1,\ldots,u_{\nu}}$. This is done by mapping the MZVs to MPLs, the MPLs to GPL using eq. \eqref{eq:mplrelation}, then applying the identity and mapping back. The requirement for the transformation to be valid, is that both MZVs are non-divergent.

{\tt Transformations["Duality"][1](m::list)} applies the duality relation eq. \eqref{eq:holder2} to $\zeta_{m_1,\ldots,m_n}$, by first mapping it to GPLs as above. Also here the only requirement is that the MZV is non-divergent.

These transformations can be accessed using the {\tt Identities} environment, as described for the GPL in section \ref{sec:implGPL}.

\subsection{PolylogTools}
\label{sec:implPolylogTools}

{\tt PolylogTools}\footnote{Please do not confuse this with a similarly named Mathematica package for symbol algebra\cite{DuhrPolyLogTools}.} is a Maple package containing functions for the manipulation of GPLs, MPLs, and MZVs. It is implemented as a member of {\tt GeneralizedPolylog} so its full name is {\tt GeneralizedPolylog:-PolylogTools}. {\tt PolylogTools} has the following public members: \\
\noindent
{\tt FindSymbol}, {\tt GetAllowNumbersInSymbols}, {\tt GetSymbolPreFactorSet},\\ {\tt IsolateDivergences}, {\tt ListShuffle}, {\tt ListStuffle}, {\tt PHC}, {\tt RemoveTrailingZeros}, {\tt SYM}, {\tt SetAllowNumbersInSymbols}, {\tt SetSymbolPreFactorSet}, {\tt SuperShuffle}, {\tt ToGPL}, {\tt ToMPL},

which will be described in the following.\\

{\tt ListShuffle(a::list, b::list)} returns $a \sha b$, i.e. the list of the shuffles of the two lists {\tt a} and {\tt b}, as defined below eq. \eqref{eq:shuffle}. The order of the members of $a \sha b$ is fixed, for instance is the first element always $\{a_1,\ldots,a_{n_a},b_1,\ldots,$ $b_{n_b}\}$. \\

{\tt ListStuffle(a::list, b::list)} returns $a \! \stuffle{\circ} \! b$, i.e. the list of the stuf- fles of the two lists {\tt a} and {\tt b}, as defined in eq. \eqref{eq:stuffleopdef}. The operator $\circ$ is not implemented directly, $\circ$ on elements $a_i$ and $b_j$ is emulated by a two-member list with elements $a_i$ and $b_j$, and then the user must apply the actual $\circ$ operator. The order of the members of $a \! \stuffle{\circ} \! b$ is fixed, such that it can be used in the stuffle product of eq. \eqref{eq:stufflepr}.\\

{\tt RemoveTrailingZeros(x)} works on an expression {\tt x} and will return the same expression, but with all GPLs in {\tt x} having their trailing zeros shuffled out using the method described in and around eq. \eqref{eq:trailingzeros}. GPLs in the output will therefore have either no trailing zeros, or have only zeros as in eq. \eqref{eq:gpldefzeros}. \\

{\tt IsolateDivergences(x)} works on an expression {\tt x}, and will return the same expression, but with all divergent GPLs having the divergence isolated as described in and around eq. \eqref{eq:isolatedivergences}. The divergent terms will all be of the form {\tt GeneralizedPolylog([x],x)}. Please note that expressions that directly contain a divergent GPL will return an error as described in section \ref{sec:implGPL}, and to avoid that, divergent GPLs have to be put in unevaluation quotes, e.g. {\tt 'GeneralizedPolylog'([x,a],x)}. \\

{\tt ToGPL(x)} works on an expression {\tt x}, and will return the same expression, but with all MPLs converted to GPLs with argument $1$, according to eq. \eqref{eq:mplrelation}. \\

{\tt ToMPL(x)} works on an expression {\tt x}, and will return the same expression, but with all GPLs converted to MPLs according to eq. \eqref{eq:mplrelation}. Please note that if GPLs with trailing zeros are present, {\tt ToMPL} returns an error. Trailing zeros have to be removed (for instance using {\tt RemoveTrailingZeros(x)}) before calling {\tt ToMPL}. \\

{\tt SuperShuffle(x,s)} works on an expression {\tt x}, and will return the same expression, but with all GPLs re-expressed using the algorithm described in section \ref{sec:numerics} (eqs. \eqref{eq:ssc1} to \eqref{eq:ssc4}) such that they have either no dependence on the constant {\tt s}, or {\tt s} as the last argument. If a GPL contain more than one index that is equal to {\tt s} the algorithm will fail, and return an error. \\

{\tt PHC(a,b)} (short for PHase Compare) is an implementation of the function $\Phi(a,b)$ defined in eq. \eqref{eq:phidef}. {\tt PHC} will be returned by {\tt SuperShuffle} but can also be used directly. \\

{\tt FindSymbol(x)} finds the ``symbol'' of an expression {\tt x} as described in section \ref{sec:symbols}. It can work on the functions {\tt GeneralizedPolylog}, {\tt MultiPolylog}, and {\tt MultiZeta}, as well as the pre-existing maple objects {\tt ln(x)}, {\tt dilog(x)}, and {\tt pi}, along with {\tt polylog(n,z)} and {\tt Zeta(n)} for {\tt n} being an integer. \\

{\tt SYM(x1,x2,...)} is used to represent the symbol as it appears in the output of {\tt FindSymbol}. The correspondence is \\[-7mm]
\begin{align}
\left(\cdots \otimes a \otimes b \otimes \cdots\right) \;\; \leftrightarrow \;\; {\tt SYM(...,a,b,...)}\,. \nn
\end{align}

{\tt SetSymbolPreFactorSet(x::set)} sets the value of an internal variable called {\tt SymbolPreFactorSet} to {\tt x}. {\tt SymbolPreFactorSet} contains the set of variables that factors outside the symbol according to eq. \eqref{eq:symrules1}, corresponding to them being treated as rational numbers. The default value of {\tt SymbolPreFactorSet} is the empty set. \\

{\tt GetSymbolPreFactorSet} returns the current value of {\tt SymbolPreFactorSet}. \\

{\tt SetAllowNumbersInSymbols(x::truefalse)} sets the value of the internal variable {\tt AllowNumbersInSymbols}. That variable determines whether or not rational numbers in the symbol tensor puts the tensor to zero or not, as discussed around eq. \eqref{eq:numberinsymbol}. The default value, {\tt true}, keeps the symbol non-zero for such cases. \\

{\tt GetAllowNumbersInSymbols} returns the current value of {\tt AllowNumbersInSymbols}. \\


\subsection{Other features}
\label{sec:implOtherFeatures}

The harmonic polylogarithm and Nielsen's polylogarithm defined in eqs. \eqref{eq:hpldef} and \eqref{eq:nielsendef} are not analytically implemented as the remaining functions defined in section \ref{sec:definitions}. Yet, an interface to the numerics for those functions, have been implemented as members of {\tt EvalfGeneralizedPolylog}: \\

{\tt EvalfGeneralizedPolylog:-evaluate{\_}HPL(a::list, x)} is a numerical implementation of the harmonic polylogarithm as defined by eq. \eqref{eq:hpldef}. Each member of {\tt a} has to belong to the set $\{0,1,-1\}$, while {\tt x} can be any complex number. {\tt evaluate{\_}HPL} is merely an interface to {\tt EvalfGeneralizedPolylog}. \\

{\tt EvalfGeneralizedPolylog:-evaluate{\_}Nielsen(n, p, z)} where {\tt n} and {\tt p} are non-negative integers, and {\tt z} is a general complex number, is a numerical implementation of Nielsen's polylogarithm as defined by eq. \eqref{eq:nielsendef}. It is merely an interface to {\tt EvalfGeneralizedPolylog}. \\

Integration of GeneralizedPolylog and MultiPolylog with the greater set of tools for manipulation of functions in Maple (such as {\tt convert} for converting to equivalent functions, {\tt diff} for differentiation, and {\tt integrate} for integration) has not been performed at the time of writing. That does not mean, however, that these features are unavailable - at least a part of that functionality is present as ``hidden exports'' of {\tt GeneralizedPolylog:-PolylogTools}. Conversion can be done using the functions {\tt ToMPL} and {\tt ToGPL} described in the previous subsection, and derivatives of GPLs with respect to their indices, are implemented as the function\\
{\tt GeneralizedPolylog:-PolylogTools:-diff{\_}of{\_}GPL(a, x, m)} which returns
\begin{align}
\frac{\partial}{\partial a_m} G(a_i,\ldots,a_m,\ldots,a_n; x) \nonumber
\end{align}
as given by eqs. \eqref{eq:gdiff1} to \eqref{eq:gdiff4}.

\section*{Final comments}

These days, the ongoing research on Feynman integrals is focusing on integrals that evaluate to functions beyond those described in this paper such as iterated integrals over elliptic integrals. (For examples, see e.g. refs. \cite{BrownLevin2011, Enriquez:2013e, Hidding:2017jkk, Broedel:2017kkb, Adams:2018yfj}.) This means that many researchers consider the generalized polylogarithms and related functions as a ``trivial'' or solved problem. Yet before a reliable implementation of that class of functions exists as a part of those mathematical tools, such as Maple, that researchers use in their daily work, this can not really be said to be the case, at least not in the author's opinion. And that is why the author hopes and believes that the implementation described in this paper will prove itself to be a useful tool for the physical and mathematical communities.\\

The Maple functions described in this paper were implemented by the author, during an internship with Maplesoft from March to May 2017. Since then the control over the code has been with Maplesoft and not with the author. This means that any bug-reports or suggestions should be directed to Maplesoft through standard channels, rather than to the author.

\section*{Acknowledgements}

The author would like to thank the HiggsTools Initial Training Network (Grant Agreement PITN-GA-2012-316704 under the European Commission) for providing the funding for the internship, during which the work described in this paper was performed. 

Specifically the author would like to thank J{\"u}rgen Gerhard and Edgardo Cheb-Terrab from Maplesoft, for advise and help during the internship. Without their support, this implementation would not have been made.

Additionally the author would like to thank Christopher Wever, Kirill Melnikov, and Kirill Kudashkin, for reading through and commenting on the manuscript in its intermediate stages.

Additional thanks go to Johannes Bl{\"u}mlein, David Broadhurst, Francis Brown, Claude Duhr, Alexander Goncharov, and Stefan Weinzierl, for replies to questions on the proper definition and domain of the multiple polylogarithm, and for the suggestion of references.

\bibliographystyle{elsarticle-num}
\bibliography{biblio.bib}

\end{document}